\documentclass[prb,preprint,groupedaddress,showpacs]{revtex4}
\usepackage{graphicx}
\begin{document}
\bibliographystyle{prbrev}

\title{Absorption spectrum of Ca atoms attached to $^4$He nanodroplets}

\author{Alberto Hernando}
\affiliation{Departament ECM, Facultat de F\'{\i}sica,
and IN$^2$UB,
Universitat de Barcelona. Diagonal 647,
08028 Barcelona, Spain}

\author{Manuel Barranco}
\affiliation{Departament ECM, Facultat de F\'{\i}sica,
and IN$^2$UB,
Universitat de Barcelona. Diagonal 647,
08028 Barcelona, Spain}

\author{Marek Kro\'snicki}
\affiliation{Institute of Theoretical Physics and Astrophysics,
University of Gdansk.
ul Wita Stwosza 57, PL80 Gdansk, Poland}

\author{Ricardo Mayol}
\affiliation{Departament ECM, Facultat de F\'{\i}sica,
and IN$^2$UB,
Universitat de Barcelona. Diagonal 647,
08028 Barcelona, Spain}

\author{Mart\'{\i} Pi}
\affiliation{Departament ECM, Facultat de F\'{\i}sica,
and IN$^2$UB,
Universitat de Barcelona. Diagonal 647,
08028 Barcelona, Spain}

\begin{abstract}

Within density functional theory,
we have obtained the structure of $^4$He droplets doped with
neutral calcium atoms.  These results
have been used, in conjunction with newly determined {\it ab-initio}
$^1\Sigma$ and $^1\Pi$ Ca-He pair potentials, to address the
$4s4p$ $^1$P$_1 \leftarrow 4s^2$ $^1$S$_0$ 
transition  of the attached Ca atom,
finding a fairly good  agreement with absorption experimental data.
We have studied the drop structure as
a function of the position of the Ca atom with respect of the center of
mass of the helium moiety. The interplay between
the density oscillations arising from the helium intrinsic structure and
the density oscillations produced by the impurity in its neighborhood 
plays a role in the determination of the
equilibrium state, and hence in the solvation properties of alkaline
earth atoms.
In a case of study, the thermal motion of the impurity within the drop 
surface region 
has been analyzed in a semi-quantitative way. We have found
that, although the atomic shift shows a sizeable
dependence on the impurity location, the thermal effect is
statistically small, contributing by about a 10\% to the
line broadening. The structure of vortices attached to the calcium
atom has been also addressed, and its effect on the calcium absorption
spectrum discussed. At variance with previous theoretical predictions,
we conclude that spectroscopic experiments on Ca
atoms attached to $^4$He drops will be likely unable to detect the
presence of quantized vortices in helium nanodrops.



\end{abstract}

\date{\today}

\maketitle

\section{Introduction}

Optical investigations of atomic impurities in superfluid helium
nanodroplets have drawn considerable attention in recent
years.\cite{Sti01,Sti06} In particular, the shifts of the electronic
transition lines with respect to the gas-phase transition lines
(atomic shifts) are a very useful observable to
determine the location of the foreign atom attached to the drop.
Alkaline earth atoms appear to play a unique role in this context.
While e.g., all alkali atoms reside in surface ``dimple'' states,
and more attractive impurities like all noble gas atoms reside in
the bulk of drops made of either isotope,\cite{Bar06} the absorption
spectra of heavy alkaline earth atoms attached to $^4$He drops
clearly support an outside location of Ca, Sr, and
Ba,\cite{Sti97,Sti99} whereas for the lighter Mg atom the
experimental evidence is that it resides in the bulk of the $^4$He
droplets.\cite{Reh00,Prz07}

We have recently presented Density Functional Theory (DFT) results
for the structure and energetics of large $^3$He and $^4$He doped
nanodroplets, showing that alkaline earth atoms from Mg to Ba go to
the bulk of $^3$He drops, whereas Ca, Sr and Ba reside in a dimple
at the surface of $^4$He drops, and  Mg is in their
interior.\cite{Her07} This is in agreement with the analysis of
available experimental data, although the case of Mg has been
questioned very recently.\cite{Ren07} Moreover, according to the
magnitude of the observed shifts, the dimple for alkaline earth
atoms was thought to be more pronounced than for alkali atoms,
indicating that the former reside deeper inside the drop than the
later. This has been also confirmed by the calculations. In
addition, the $5s5p\leftarrow5s^2$ experimental transition of Sr
atoms attached to helium nanodroplets of either isotope has shown
that strontium is solvated inside $^3$He nanodroplets, also
in agreement with the calculations.\cite{Her07}

Calcium atoms are barely stable on the surface of the drop, and the
difference between the energy of the surface dimple state and that
of the solvated state in the bulk of the drop is rather small and
depends very sensitively on the Ca-He interatomic
potential.\cite{Anc03b} The aim of this work is to obtain the atomic
shifts for Ca attached to large $^4$He drops, and to compare them
with the experimental data. To this end, we have improved our DFT
approach,\cite{Her07} treating the atomic impurity as a quantal
particle instead of as an external field.  Laser Induced
Fluorescence (LIF) experiments for Ca atoms in liquid $^3$He and
$^4$He have been reported\cite{Mor05} and analyzed within a
vibrating bubble model, which involves the formation of a bubble
around the impurity, using Ca-He pair potentials based on
pseudopotential SCF/CI calculations.\cite{Czu91}

This work is organized as follows. In Sec. II we discuss the Ca-He
interaction potentials we have used. In Sec. III we briefly present
our density functional approach, as well as some illustrative
results for the structure of Ca@$^4$He$_N$ drops. The method we have
employed to obtain the atomic shifts is discussed in Sec. IV. In
Sec. V we present the results obtained for calcium, discuss how
thermal motion may affect the line shapes, and investigate how the
presence of a quantized vortex line may change the Ca absorption
spectrum. Finally, a summary is presented in Sec. VI.

\section{Calcium-helium interaction potentials}

Figure \ref{fig1} shows the $X^1\Sigma$ Ca-He adiabatic potential
obtained by different authors.\cite{Meyer,Par01,Hin03,Lov04} This
potential determines the dimple structure described in the
next Section. It can be seen that apart from the unpublished
potential by Meyer, the others are quite similar. As in our previous
work,\cite{Her07} we shall use the one obtained in Ref.
\onlinecite{Lov04}. This will allow us to ascertain the effect of
treating Ca as a quantal particle. Since the $X^1\Sigma$ potential
seems to be fairly well determined, we have turned our attention to
the excited adiabatic potentials.

In a previous work\cite{Czu03} the excited state potentials were
calculated in  a valence \emph{ab-initio} scheme. The core electrons
of calcium and helium were replaced by scalar-relativistic
energy-consistent pseudopotentials, and the  energy curves were
calculated  on the  complete-active-space multiconfiguration
selfconsisted field (CASSCF)\cite{WER85,KNO85}/complete-active-space
multireference second order perturbation level of theory.

In this work we have performed fully \emph{ab-initio} calculations
and have only focused on singlet states. The calculations for
excited states have been done at the CASSCF/internally contracted
multireference configuration interaction (ICMRCI)\cite{WER88,KNO88}
level of theory. In the calculations we have  used correlation
consistent polarized valence five zeta (cc-pV5Z) basis sets. For the
calcium atom we have used the
(26s,18p,8d,3f,2g,1h)/[8s,7p,5d,3f,2g,1h] basis set developed  by
Koput and  Peterson,\cite{JKKP02} and for the helium atom we have
used the (8s,4p,3d,2f,1g)/[5s,4p,3d,2f,1g]  basis set  developed by
Woon and Dunning.\cite{DWTD94}

The calculations were performed within the MOLPRO suite of
\emph{ab-initio} programs.\cite{MOLPRO} The molecular orbitals used
for the excited states  calculations were optimized in the state
averaged CASSCF method for all singlet states correlating to
(4s$^2$)$^1$S, (4s3d)$^1$D, and (4s4p)$^1$P atomic asymptotes. The
active space was formed by distributing the two valence electrons of
the Ca atom into $4s3d4p$  valence orbitals. The  $1s2s2p3s3p$
orbitals of calcium and  $1s$ orbital of helium were kept doubly
occupied in all configuration state functions, but they were
optimized in the CASSCF calculations.  The resulting wave functions
were used as  references in the following ICMRCI calculations. At
the ICMRCI level, the $1s2s2p3s3p$ orbitals of calcium and  $1s$
orbital of helium  were kept  as doubly occupied in all reference
configuration state functions, but these orbitals were correlated
through single and double excitations. The  $4s$, $4p$, and $3d$
calcium  orbitals had not restricted occupation patterns. We show in
Fig. \ref{fig2} the excited adiabatic potentials; to have better
insight into the potential minima of the $^1\Sigma$ and $^1\Pi$
potentials, they have been plotted correlating to the (4p)$^1$P Ca
term.

\section{DFT description of helium nanodroplets}

In recent years, static and time-dependent density functional
methods\cite{Dal95,Gia03,Leh04} have become increasingly popular to
study inhomogeneous liquid helium systems because they provide an
excellent compromise between accuracy and computational effort,
allowing to address problems inaccessible to more fundamental
approaches, see e.g. Ref. \onlinecite{Bar06} for a recent review.
Obviously, DFT cannot take into account the atomic, discrete nature
of these systems, but can address inhomogeneous helium systems at the
nanoscale\cite{Anc05} and take into account the anisotropic
deformations induced by some dopants in helium drops. Both
properties are essential to properly describe these systems.

Our starting point is the Orsay-Trento density
functional,\cite{Dal95} together with the Ca-He adiabatic potential
$X^1\Sigma$ of Ref. \onlinecite{Lov04}, here denoted as $V_{Ca-He}$.
This allows us to write the energy of the Ca-drop system as a
functional of the Ca wave function $\Phi(\mathbf{r})$ and the $^4$He
``order parameter'' $\Psi(\mathbf{r})$:

\begin{eqnarray}
E[\Psi, \Phi] &=&
\frac{\hbar^2}{2\,m_{He}} \int \mathrm{d}^3 \mathbf{r}\, |\nabla \Psi(\mathbf{r}\,)|^2 +
\int \mathrm{d}^3 \mathbf{r} \, {\cal E}(\rho)
\nonumber \\
&+ &\frac{\hbar^2}{2\,m_{Ca}}\int \mathrm{d}^3 \mathbf{r}\, |\nabla \Phi(\mathbf{r}\,)|^2
+ \int \int \mathrm{d}^3 \mathbf{r}\,  \mathrm{d}^3 \mathbf{r'}\,
|\Phi(\mathbf{r})|^2 \, V_{Ca-He}(|\mathbf{r}-\mathbf{r'}|)\,\rho(\mathbf{r'})  \; .
\label{eq1}
\end{eqnarray}
The order parameter is defined as $\Psi(\mathbf{r}) = \sqrt{\rho(\mathbf{r})}
\exp [\imath S(\mathbf{r})]$, where $\rho(\mathbf{r})$ is the particle
density and $\mathbf{v}(\mathbf{r}) = \hbar \nabla
S(\mathbf{r})/m_4$ is the velocity field of the superfluid.
In Eq. (\ref{eq1}), ${\cal E}(\rho)$ is the $^4$He ``potential
energy density''.\cite{Dal95} In the absence of vortex lines, we 
set $S$ to zero and $E$ becomes a functional of $\rho$ and $\Phi$.
Otherwise, we have used the complex order parameter $\Psi(\mathbf{r})$
to describe the superfluid.

We have solved the Euler-Lagrange equations which result from the
variations with respect to $\Psi^*$ and $\Phi^*$ of the energy
$E[\Psi, \Phi]$ under the constrain of a given number of helium
atoms in the drop, and a normalized Ca wave function, namely:

\begin{equation}
-\frac{\hbar^2}{2\, m_{He}}\Delta \Psi +
\left\{\frac{\delta {\cal E}}{\delta \rho}
+U_{He} \right\} \Psi = \mu \Psi
\label{eq2}
\end{equation}
\begin{equation}
-\frac{\hbar^2}{2\, m_{Ca}}\Delta \Phi
+U_{Ca}\, \Phi  =   \varepsilon \Phi   \; ,
\label{eq3}
\end{equation}
where $\mu$ is the helium chemical potential and $\varepsilon$ is
the lowest eigenvalue of the Schr\"odinger equation obeyed by the Ca
atom. The effective potentials $U_{He}$ and $U_{Ca}$ are defined as

\begin{eqnarray}
U_{He}(\mathbf{r})&=&
\int \mathrm{d}^3 \mathbf{r'}\, |\Phi(\mathbf{r'})|^2 
\, V_{Ca-He}(|\mathbf{r}-\mathbf{r'}|)
\nonumber \\
U_{Ca}(\mathbf{r})&=&
\int \mathrm{d}^3 \mathbf{r'}\, \rho(\mathbf{r'})  
\, V_{Ca-He}(|\mathbf{r}-\mathbf{r'}|)
\label{eq4} \; .
\end{eqnarray}
The coupled Eqs. (\ref{eq2}-\ref{eq3}) have to be solved
selfconsistently, starting from an arbitrary but reasonable choice
of the unknown functions $\Psi$ and $\Phi$. 
In spite of the axial symmetry of the problem, we have solved them
in three-dimensional (3D) cartesian coordinates. The
main reason is that these coordinates allow us to use fast Fourier
transformation techniques\cite{FFT} to efficiently compute the
convolution integrals entering the definition of ${\cal E}(\rho)$,
i.e. the mean field helium potential and the coarse-grained density
needed to compute the correlation term in the He density functional,
\cite{Dal95} as well as the fields defined in Eqs. (\ref{eq4}).

The differential operators in Eqs. (\ref{eq1}-\ref{eq3}) have been
discretized  using 13-point formulas for the derivatives, and Eqs.
(\ref{eq2}-\ref{eq3}) have been solved employing an imaginary time
method;\cite{Pre92} some technical details of our procedure are
given in Ref. \onlinecite{Anc03a}. Typical calculations have been
performed using a spatial mesh step of 0.5 \AA. We have checked the
stability of the solutions against reasonable changes in the step.

Equations (\ref{eq2}-\ref{eq3}) have been solved for several $N$
values from 100 to 2000. They will allow us to study the atomic
shift as a function of the cluster size. The equilibrium
configurations of  Ca@$^4$He$_{1000}$ and  Ca@$^4$He$_{2000}$ will
be shown later on.

Figure \ref{fig3} shows the energy of a Ca atom attached to a drop,
defined as the energy difference

\begin{equation}
S_N(Ca)= E(Ca@^4He_N) - E(^4He_N) \; .
\label{eq5}
\end{equation}
On the figure are shown also the results obtained treating calcium as an 
external field.\cite{Her07} It can be seen that for large drops the energy 
of the calcium atom is about 10 K less negative due to its zero point
motion. We want to stress again how barely stable is the calcium
atom on the surface of $^4$He$_N$ drops. For instance, we have found
that the total energy of the equilibrium -dimple- configuration of
Ca@$^4$He$_{1000}$ is $\sim -5467.4$ K, whereas it is $\sim -5455.0$
K when Ca is forced to be at the center of the drop. For
Ca@$^4$He$_{500}$, the corresponding values are 
$\sim -2525.2$ K and $\sim -2511.1$ K, respectively.

The dimple depth $\xi$, defined as the difference between the
position of the dividing surface at $\rho=\rho_0/2$, where
$\rho_0=0.0218$ \AA$^{-3}$ is the bulk liquid density, with and
without impurity, respectively, is shown in Fig. \ref{fig4}. Due to
the zero point motion that pushes the impurity towards lower helium
densities, for large drops the dimple depth is about 0.8 \AA{}
smaller when the zero point motion is included than when it is
not.\cite{Her07} This change in the depth is large enough to produce
observable effects in the calculated absorption spectrum, as
discussed below.

It can be seen that the dimple depth curve $\xi(N)$ has some
structure. This is not a numerical artifact, but a genuine effect
due to the interplay between the Ca atom and the drop, whose
density, even for pure drops, shows conspicuous oscillations all
over the drop volume, extending up to the surface region
irrespective of whether the drop is described within DFT or
Diffusion Monte Carlo methods.\cite{Bar06,Dal95,Chi92} The interplay
of these oscillations with those arising from the presence of the
impurity little affects the total energy of the system and hence,
the Ca energy, but yields some visible structure in the density
distributions that shows up in related quantities, like the dimple
depth. To illustrate it, we show in Fig. \ref{fig5} the density of
the helium moiety of Ca@$^4$He$_{2000}$, where the interference
pattern can be clearly seen.

Further insight can be gained studying, for a given drop, the energy
of the Ca@$^4$He$_N$ complex as a function of the distance between
the centers of mass of the impurity and of the helium moiety. This
can be done adding an appropriate constraint to the total energy in
Eq. (\ref{eq1}), and solving the corresponding Euler-Lagrange
equations. Specifically, we have minimized the expression

\begin{equation}
E + \frac{\lambda_C}{2} [{\cal Z} - {\cal Z}_0]^2 \; ,
\label{eq3bis}
\end{equation}
where ${\cal Z}$ is the average distance in the $z$ direction
between the impurity and the geometrical center of the helium moiety

\begin{equation}
{ \cal Z} =
\int d \mathbf{r}^3 \, z \, |\Phi(\mathbf{r})|^2  -
\frac{1}{N} \int d \mathbf{r}^3 \, z \,\rho(\mathbf{r})
\; ,
\label{eq4bis}
\end{equation}
and $\lambda_C$ is an arbitrary constant, large enough to guarantee
that upon minimization, ${\cal Z}$ equals the desired ${\cal Z}_0$
value. 
We have also applied this method to Mg doped helium drops, and
will present the details of the calculation elsewhere.\cite{Her07b}

We show in the bottom panel of Fig. \ref{fig6} the total energy of
the Ca@$^4$He$_{500}$ and Ca@$^4$He$_{1000}$ systems as a function
of ${\cal Z}_0$.  For the sake of comparison, the energies and
${\cal Z}_0$ distances are referred to their equilibrium values. The
vertical lines roughly delimit the drop surface regions,
conventionally defined as the radial distance between the points
where the density equals 0.1$\rho_0$ and 0.9$\rho_0$ (see Fig.
\ref{fig10}). The horizontal line has been drawn 0.4 K above the
equilibrium energy, representing the accessible energy range
due to the temperature of the helium drops.\cite{Toe04} Its
intersection with the energy curves yields a qualitative measure of
the dispersion of the impurity location due to
thermal motion. The energy curve of Ca@$^4$He$_{1000}$
displays some structure to the left of the minimum due to
the mentioned interference pattern. 
This behavior has not been disclosed before, and appears in the
course of ``pushing'' the impurity inside the droplet from its
equilibrium position (a similar structure shows up for
Ca@$^4$He$_{500}$, but at more negative $\Delta {\cal Z}_0$ values).
While it affects
rather little the equilibrium location of the Ca atom because of its
clear surface location, and hence the atomic shift -see however
the inset in Fig. \ref{fig8}, it plays
a substantial role in the solvation of magnesium atoms in small helium
drops.\cite{Her07b}
We recall that this problem has recently drawn the attention of
experimentalists and theoreticians as 
well.\cite{Reh00,Prz07,Her07,Ren07,Mel05,Elh07}
How the position of the Ca atom affects the absorption spectrum
throughout the change in the dimple structure
will be discussed in Sec. V.

\section{Excitation spectrum of an atomic impurity in a $^4$He drop}

Lax method\cite{Lax52} offers a realistic way to study the
absorption spectrum $I(\omega)$ of a foreign atom embedded in helium
drops. It makes use of the Franck-Condon principle within a
semiclassical approach, and it has been employed to study the
absorption spectrum of several atomic dopants attached to fairly
small $^4$He drops.\cite{Che96,Nak01,Mel02,Mel05} The case of alkali
atoms attached to large drops described within DFT has been also
considered, see Refs. \onlinecite{Sti96,Bue07} and references
therein. Lax method is usually applied in conjunction with the
diatomics-in-molecules theory,\cite{Ell63} which means that the
atom-drop complex is treated as a diatomic molecule, where the
helium moiety plays a role of the other atom.

In the original formalism, to obtain $I(\omega)$ one has to carry
out an average on the possible initial states of the system that may
be thermally populated. Usually, this average is not needed for
helium drops, as their temperature, about $0.4$ K,\cite{Toe04} is
much smaller than the vibrational excitation energies of the Ca atom
in the mean field represented by the second of 
Eqs. (\ref{eq4}).\cite{Note0}
However, thermal broadering due to the ``wandering'' of the dopant
must be analyzed separately if it is in a dimple state, as this may
have some influence on the line shape. Its effect on the absorption
spectrum will be exemplified later on for the $N=500$ case.
Contrarily, when the impurity is very attractive and resides in the
bulk of the drop, thermal motion plays no role, as the dopant hardly
gets close enough to the drop surface to have some effect on the
line shape. In this case, dynamical deformations of the cavity
around the impurity may be relevant (Jahn-Teller effect). 
It cannot be discarded that, if some of these very attractive impurities
have an angular momentum large enough,\cite{Lehm04} they may get
close to the drop surface, in which case thermal motion might have some
influence on the absorption spectrum.

We review here the essentials of the method and the way we have
implemented it. In particular, we present some of the expressions in
cartesian coordinates, better adapted to our approach. They are of
course equivalent to the expressions in spherical coordinates that
can be found in the literature -see e.g. Ref. \onlinecite{Nak01} and
references therein.

\subsection{Line shapes}

The line shapes for electronic
transitions from the ground state $(gs)$ to the excited state $(ex)$
in a condensed phase system can be written as

\begin{equation}
I(\omega)\propto
\int\mathrm{d}t~\mathrm{e}^{-i\omega t}\langle \Psi^{gs}| D_{ge}^\dag
~\mathrm{e}^{\frac{i t}{\hbar}\mathcal{H}_{ex}}~D_{ge}~\mathrm{e}^{-\frac{i t}{\hbar}
\mathcal{H}_{gs}} | \Psi^{gs} \rangle   \; ,
\label{e1}
\end{equation}
where $D_{\mathrm{ge}}$ is the matrix element of the electric dipole
operator, $\mathcal{H}_{gs}$ and $\mathcal{H}_{ex}$ are the
Hamiltonians which describe the ground and excited states of the
system respectively, and $|\Psi^{gs}\rangle$ represents the ground
state.
$I(\omega)$ can be evaluated using the Born-Oppenheimer
approximation, which makes a separation of the electronic and
nuclear wave functions $|\Psi^i\rangle=|e^i\rangle |\psi^i\rangle$,
and the Franck-Condon principle, whereby the heavy nuclei do not
change their positions or momenta during the electronic transition.
If the excited electron belongs to the impurity, the helium cluster
remains frozen, so that the relevant coordinate is the relative
position $\mathbf{r}$ between the cluster and the impurity. That
principle amounts to assuming that $D_{\mathrm{ge}}$ is independent
of the nuclear coordinates. Taking into account that
$\mathrm{e}^{-\frac{it}{\hbar}\mathcal{H}_{gs}}|\Psi^{gs}\rangle=
\mathrm{e}^{-it \omega_{gs}} |\Psi^{gs}\rangle$ and projecting on
eigenstates of the orbital angular momentum of the excited electron
$| m \rangle$,

\begin{eqnarray}
I(\omega)&\propto&\sum_{m}\int\mathrm{d}t~\mathrm{e}^{-i(\omega+\omega^{\mathrm{gs}})t}
\langle \Psi^{\mathrm{gs}}|D_{\mathrm{ge}}^{\dag} |m\rangle~\mathrm{e}^{-\frac{i t}
{\hbar}\mathcal{H}^{\mathrm{ex}}_m}\langle m |D_{\mathrm{ge}} |\Psi^{\mathrm{gs}}\rangle
\nonumber
\\
&=&\sum_{m}
\int\mathrm{d}t~\mathrm{e}^{-i(\omega+\omega^{\mathrm{gs}})t}
\int\mathrm{d}^3\mathbf{r}\int\mathrm{d}^3\mathbf{r'}~
\langle \Psi^{\mathrm{gs}}|\mathbf{r}\rangle \langle \mathbf{r}|D_{\mathrm{ge}}^{\dag}
|m \rangle\mathrm{e}^{\frac{i t}
{\hbar}\mathcal{H}^{\mathrm{ex}}_m}\langle m |D_{\mathrm{ge}}|\mathbf{r'}\rangle
\langle\mathbf{r'}|\Psi^{\mathrm{gs}}\rangle
\nonumber
\\
&=&|D_{\mathrm{ge}}|^2\sum_{m}\int\mathrm{d}t~\mathrm{e}^{-i(\omega+\omega^{\mathrm{gs}}_X)t}
\int\mathrm{d}^3\mathbf{r}~
{\psi_X^{\mathrm{gs}}(\mathbf{r})}^*~\mathrm{e}^{\frac{i t}
{\hbar}H^{\mathrm{ex}}_m(\mathbf{r})}~\psi_X^{\mathrm{gs}}(\mathbf{r}) \; ,
\label{fourier}
\end{eqnarray}
where $\hbar\omega^{\mathrm{gs}}_X$ and
$\psi_X^{\mathrm{gs}}(\mathbf{r})$ are the energy and the wave
function of the ro-vibrational ground state of the frozen
helium-impurity system, and $H^{\mathrm{ex}}_m(\mathbf{r})$ is the
ro-vibrational excited Hamiltonian with  potential energy
$V^{\mathrm{ex}}_m(\mathbf{r})$ determined by the electronic energy
eigenvalue, as obtained in the next subsection for a $p \leftarrow
s$ transition. Eq. (\ref{fourier}) will be referred to as the
\emph{Fourier Formula}, and it is the Fourier transform of the
time-correlation function. It is nothing but a sum at the resonant
energies weighted with the well-known Franck-Condon factors:

\begin{eqnarray}
I(\omega)&\propto&\sum_{m}\int\mathrm{d}t~\mathrm{e}^{-i(\omega+\omega^{\mathrm{gs}}_X)t}
\langle \psi_X^{\mathrm{gs}}|~
\mathrm{e}^{\frac{i t}{\hbar}H^{\mathrm{ex}}_m}~|\psi_X^{\mathrm{gs}}\rangle
\nonumber
\\
&=&\sum_{m}\sum_{\nu~\nu'}\int\mathrm{d}t~\mathrm{e}^{-i(\omega+\omega^{\mathrm{gs}}_X)t}
\langle\psi_X^{\mathrm{gs}}|\psi^m_\nu\rangle \langle \psi^m_\nu|~\mathrm{e}^{\frac{i t}
{\hbar}H^{\mathrm{ex}}_m}~|\psi^m_{\nu'}\rangle\langle\psi^m_{\nu'}|\psi_X^{\mathrm{gs}}\rangle
\nonumber
\\
&=&\sum_{m}\sum_{\nu}\int\mathrm{d}t~\mathrm{e}^{-i(\omega+\omega^{\mathrm{gs}}_X-
\omega^m_\nu)t}|\langle\psi^m_\nu|\psi_X^{\mathrm{gs}}\rangle|^2
\nonumber
\\
&=&\sum_{m}\sum_{\nu}\delta(\omega+\omega^{\mathrm{gs}}_X-\omega^m_\nu)|
\langle\psi^m_\nu|\psi_X^{\mathrm{gs}}\rangle|^2  \; ,
\nonumber
\end{eqnarray}
where $\omega^m_\nu$ and $|\psi^m_\nu\rangle$ are the ro-vibrational
eigenvalues and eigenstates of the Hamiltonian $H^{\mathrm{ex}}_m$.

If the relevant excited states for the transition have large quantum
numbers, they can be treated as approximately 
classical\cite{Lax52,Che96,Nak01} using the
averaged energy $\hbar\omega^m_\nu\approx
V^{\mathrm{ex}}_m(\mathbf{r})$ which is independent of $\nu$. In
this case we obtain the expression

\begin{eqnarray}
I(\omega)&\propto&
\sum_{m}\int\mathrm{d}^3\mathbf{r}~|\psi_X^{\mathrm{gs}}(\mathbf{r})|^2
\delta(\omega+\omega^{\mathrm{gs}}_X-V^{\mathrm{ex}}_m(\mathbf{r})/\hbar)
\nonumber
\\
&=& \hbar \int_{\Omega_m(\omega)} \;
\mathrm{d}^2\mathbf{r}~\frac{|\psi_X^{\mathrm{gs}}(\mathbf{r})|^2}
{|\mathbf{\nabla}V^{\mathrm{ex}}_m(\mathbf{r})|}   \, ,
\label{sc}
\end{eqnarray}
where $\Omega_m(\omega)$ is the surface defined by the equation
$\omega+\omega^{\mathrm{gs}}_X-V^{\mathrm{ex}}_m(\mathbf{r})/\hbar=0$.
We will refer to this equation as the \emph{Semi-Classical Formula}.

If the atom is in bulk liquid helium, or at the center of the drop,
the problem has spherical symmetry and the above equation reduces to

\begin{eqnarray}
I(\omega)&\propto&
4\pi\sum_{m}\int\mathrm{d}r~|r~\psi_X^{\mathrm{gs}}(r)|^2
\delta(\omega+\omega^{\mathrm{gs}}_X-V^{\mathrm{ex}}_m(r)/\hbar)
\nonumber
\\
&=& 4\pi\hbar\sum_{m}\left|\frac{[r~\psi_X^{\mathrm{gs}}(r)]^2}
{d{V}^{~\mathrm{ex}}_m(r)/d r}\right|_{r=r_m(\omega)}
\; ,
\label{scsc}
\end{eqnarray}
where $r_m(\omega)$ is the root of the equation
$\omega+\omega^{\mathrm{gs}}_X-V^{\mathrm{ex}}_m(r)/\hbar=0$.

In the non-spherical case, we have evaluated $I(\omega)$ from the
first expression in Eq. (\ref{sc}) using the discretization

\begin{eqnarray}
I(n\Delta\omega+\omega_0)&\propto&
\sum_{m}\sum_{ijk}|\psi_X^{\mathrm{gs}}(\Delta\mathbf{r}_{ijk}+\mathbf{r}_0)|^2
\frac{\Delta x\Delta y\Delta z}{\Delta \omega}
\nonumber
\\
&\times&\left\{\Theta\left[\left(n-1/2\right)\Delta\omega+\omega_0+\omega^{\mathrm{gs}}_X
-V^{\mathrm{ex}}_m(\Delta\mathbf{r}_{ijk}+\mathbf{r}_0)/\hbar\right]\right.
\nonumber
\\
&-&\left.\Theta\left[\left(n+1/2\right)\Delta\omega+\omega_0+\omega^{\mathrm{gs}}_X-
V^{\mathrm{ex}}_m(\Delta\mathbf{r}_{ijk}+\mathbf{r}_0)/\hbar\right]\right\}
\; ,
\end{eqnarray}
where $\Theta$ is the step function and $\Delta \omega$ is a
frequency step small enough so that the above discretization
represents the delta function. We also take advantage that only
points near the impurity contribute to the integral, by writing
$\Delta\mathbf{r}_{ijk}+\mathbf{r}_0=(i\Delta x+x_0,j\Delta
y+y_0,k\Delta z+z_0)$, with $\Delta x$, $\Delta y$, $\Delta z$ being
the spatial mesh steps used in the discretization, and being
$\mathbf{r_0}$ an arbitrary point in the neighborhood of the
impurity. Finally, we recall that $I(\omega)$ needs to be  evaluated
only in a narrow frequency range starting from an arbitrary
$\omega_0$ which can be, e.g., the free atom frequency. This range
defines the  maximum $n$ value in the above equation.

\subsection{Excited ro-vibrational potential for a
$\mathbf {p \leftarrow s}$ transition}

We next determine the potential energy surfaces
$V^{\mathrm{ex}}_m(\mathbf{r})$ needed to carry out the calculation
of the atomic shifts.

\subsubsection{Pairwise sum aproximation}
The pair-interaction between an atom in a $s$-state and an atom in a
$p$-state can be expressed, in the cartesian eigenbasis
($|x\rangle,|y\rangle,|z\rangle$) as

\begin{eqnarray}
U(r)&=&
        \left(
\begin{array}{ccc}
       V_{\Pi}(r) & 0 & 0 \\
       0 & V_{\Pi}(r) & 0 \\
       0 & 0 & V_{\Sigma}(r)
\end{array}
        \right)
\nonumber
        \\
&=&V_{\Pi}(r)\{|x\rangle\langle x| +|y\rangle\langle y|\} +V_{\Sigma}(r)|z\rangle\langle z|
\nonumber
\\
&=&V_{\Pi}(r)\mathbf{I}+\{V_{\Sigma}(r)-V_{\Pi}(r)\}|z\rangle\langle z |  \; ,
\label{pairwise}
\end{eqnarray}
where $V_\Pi(r)$ and $V_\Sigma(r)$ are the adiabatic potentials neglecting
the spin-orbit interaction and $r$ is the distance between atoms.
For a system of $N$ helium atoms and an excited impurity in a
$p$-state, the total potential is approximated by the pairwise sum

\begin{equation}
U=\sum_{n=1}^N\left\{V_\Pi(r_n)\mathbf{I}+[V_\Sigma(r_n)-
V_\Pi(r_n)]R_n|z\rangle\langle z |R^{-1}_n\right\}  \; ,
\end{equation}
where $r_n$ is the distance between the $n^{\mathrm{th}}$ helium
atom and the impurity, and $R_n$ is the rotation matrix that
transform the unity vector $\mathbf{\hat{z}}$ into the
$\mathbf{\hat{r}}_n$ vector. It can be shown that, in cartesian
coordinates,

\begin{equation}
\langle x^i| R_n|z\rangle\langle z| R^{-1}_n|x^j\rangle=\frac{x_n^i~x_n^j}{r_n^2}\; ,
\end{equation}
where $x^1=x$, $x^2=y$, $x^3=z$, and $r_n^2=x_n^2+y_n^2+z_n^2$.
Thus, the matrix elements of the total potential are

\begin{equation}
\langle x^i| U|x^j\rangle\equiv U_{ij}=
\sum_{n=1}^N\left\{V_\Pi(r_n)\delta_{ij}+
[V_\Sigma(r_n)-V_\Pi(r_n)]\frac{x_n^i~x_n^j}{r_n^2}\right\}  \; .
\end{equation}
Using the continuous density approach inherent to DFT
$[\sum_n\rightarrow\int\mathrm{d}^3\mathbf{r'}\rho(\mathbf{r'})]$,
this expression can be written as

\begin{equation}
U_{ij}(\mathbf{r})= \int\mathrm{d}^3\mathbf{r'}\rho(\mathbf{r'}+\mathbf{r})
\left\{V_\Pi(r')\delta_{ij}+[V_\Sigma(r')-V_\Pi(r')]\frac{x'^i~x'^j}{r'^2}\right\} \; .
\label{v}
\end{equation}
The eigenvalues of this symmetric matrix are the sought-after
$V^{\mathrm{ex}}_m(\mathbf{r})$ which define the potential energy
surfaces (PES) as a function of the distance between the centers of
mass of the droplet and of the impurity, and are given by the three
real roots $\lambda_i(\mathbf{r})$ of the equation

\begin{equation}
\lambda^3+C \lambda^2+B \lambda +A=0
\end{equation}
with
\begin{eqnarray}
C&=&-\sum_{i=1}^3U_{ii}
\nonumber
\\
B&=&\frac{1}{2}\sum_{i\neq j}^3\left(U_{ii}U_{jj}-U^2_{ij}\right)
\nonumber
\\
A&=&\sum_{i\neq j\neq k}^3
\left(\frac{1}{2}U_{ii}U_{jk}^2-\frac{1}{3}U_{ii}U_{jj}U_{kk}
-\frac{2}{3}U_{ij}U_{jk}U_{ki}\right)
\; .
\label{abc}
\end{eqnarray}

It can be shown that for spherical geometry, Eq. (\ref{v}) is
diagonal with matrix elements (in spherical coordinates)
\begin{eqnarray}
\lambda_i(r) \equiv
U_{ii}(r)&=&2\pi\int\int
r'^2\sin\theta'\mathrm{d}\theta'\mathrm{d}r'\rho(|r'+r+2r'r\cos\theta'|)
\nonumber
\\
&\times&\left\{V_\Pi(r')+[V_\Sigma(r')-V_\Pi(r')]
\left[\frac{1}{2}(\delta_{i1}+\delta_{i2}) \sin^2\theta'
+\delta_{i3} \cos^2\theta'
\right]\right\}
\label{spherical}
\end{eqnarray}

\subsubsection{Spin-Orbit coupling}

For atomic impurities in which the spin-orbit (SO) interaction is
prominent and comparable to the splitting of the $P$-states due to
the interaction with the droplet, it has to be taken into account in
the calculation of the PESs. This is usually done considering that
the SO splitting of the dopant is that of the isolated atom
irrespective of the impurity-drop distance.\cite{Nak01,Coh74,Jak97}
Given the atomic structure of alkaline earth atoms, the SO
interaction can be safely neglected in the PES calculation. However,
we discuss it here for the sake of completeness and future
reference.

When the spin-orbit interaction is taken into account, the total
potential can be written as $V_T=U+V_{SO}$, where $V_{SO}$ has the
form, in the spin-cartesian orbit basis
(${|x,1/2\rangle,|x,-1/2\rangle,|y,1/2\rangle,|y,-1/2\rangle,|z,1/2\rangle,|z,-1/2\rangle}$):

\begin{eqnarray}
V_{SO}&=& \frac{A_{\ell s}}{2}
        \left(
\begin{array}{cccccc}
         0 & 0 & -i & 0 & 0 & 1 \\
         0 & 0 & 0 & i & -1 & 0 \\
         i & 0 & 0 & 0 & 0 & -i \\
         0 & -i & 0 & 0 & -i & 0 \\
         0 & -1 & 0 & i & 0 & 0 \\
         1 & 0 & i & 0 & 0 & 0
\end{array}
        \right)   \; ,
\nonumber
\end{eqnarray}
where $A_{\ell s}$ is $2/3$ of the experimental SO splitting of an
isolated atom. Kramers' theorem states that there is a two-fold
degenerate manyfold of systems with a total half-integer spin value
that cannot be broken by electrostatic interactions,\cite{Mei62} so
that the two-fold degenerate eigenvalues that define the PES's are
the roots of the equation

\begin{equation}
\lambda^3+C \lambda^2+[B-\frac{3}{4}A_{\ell s}^2] \lambda +
\{ A +\frac{1}{4}[A_{\ell s}^3-A_{\ell s}^2 C]\}=0
\end{equation}
with $A$, $B$ and $C$ defined in Eq. (\ref{abc}). In the case of
spherical geometry, the eigenvalues adopt a simple expression:

\begin{eqnarray}
\lambda_1(r)&=&\frac{1}{2}(U_{11}+U_{33})+\frac{1}{4}\left[-A_{\ell s}
+\sqrt{9A_{\ell s}^2-4A_{\ell s}(U_{11}-U_{33})+4(U_{11}-U_{33})^2}\right]
\nonumber
\\
\lambda_2(r)&=&U_{11}+\frac{A_{\ell s}}{2}
\nonumber
\\
\lambda_3(r)&=&\frac{1}{2}(U_{11}+U_{33})+\frac{1}{4}\left[-A_{\ell s}
-\sqrt{9A_{\ell s}^2-4A_{\ell s}(U_{11}-U_{33})+4(U_{11}-U_{33})^2}\right]
\; ,
\end{eqnarray}
which reduces to Eq. (\ref{spherical}) if  $A_{\ell s}=0$. 

It is important to notice that for spherical geometries (spherically
symmetric impurities in liquid helium or at the center of a drop),
when the SO interaction is negligible two of the PES are degenerate,
as it can be seen from Eq. (\ref{spherical}), that yields
$\lambda_1(r)=\lambda_2(r)\neq\lambda_3(r)$.\cite{Note1}
Thus, the existence of the SO interaction not only is the reason of
the appearance of the $D_1$ and $D_2$ lines in the case, e.g.,  of
alkali atoms in bulk liquid helium, but also the reason of either
the broadening or the splitting of the $D_2$ line.\cite{Kin95} For
this particular geometry, another contribution to the splitting of
the $D_2$ line is the Jahn-Teller effect caused by dynamical quadrupole
deformations of the cavity surrounding the impurity, which develop
irrespective of whether the spin-orbit energy is relevant or
not.\cite{Reh00,Kin96} When the impurity resides in a deformed
environment like a dimple, the three PES are non degenerate, and may
cause the appearance of three distinct peaks in the absorption
spectrum, or of just one single broad peak, as it happens in the
case of Ca, Sr and Ba atoms attached to $^4$He
drops.\cite{Sti97,Sti99} The liquid $^4$He results for alkaline
earth atoms are reported in Refs. \onlinecite{Bau90,Mor06} and
references therein.

\section{Results for the absorption spectrum of calcium atoms}

\subsection{Line shifts}

The problem of obtaining the line shifts has been thus reduced to
that of the dopant in the 3D trapping potentials corresponding to
the ground state, $U_{Ca}(\mathbf{r})$, and $P$ excited states,
$\lambda_i(\mathbf{r})$. Since we have neglected the fluctuations of
the dimple -shape fluctuations\cite{Ler93}- and their coupling to
the dopant dipole oscillations, as well as inhomogeneous broadening
resulting from droplet size distributions, laser line width and
similar effects, the model is not expected to yield the line shapes,
but only the energies of the atomic transitions. These limitations
are often overcome by introducing line shape functions or
convoluting the calculated lines with some effective line
profiles.\cite{Sti96,Bue07} We discuss now some illustrative
examples without considering these justified but somewhat
uncontrolled convolutions.

Figure \ref{fig7} shows the absorption spectrum of
Ca@$^4$He$_{2000}$ calculated with the the Semi-Classical Formula.
The much involved Fourier Formula calculation is unnecessary in the
Ca@$^4$He$_N$ case. The reason is the absence of bound-bound
transitions from the ground state PES to the $\Pi$ or $\Sigma$ ones
because their wells are spatially well apart. The starred
vertical line represents the gas-phase transition. The three
components of the absorption line, each arising from a different
excited PES, are also shown. We have normalized to one the integral
of each component. This choice comes out naturally from the
normalization of the wave function of the impurity; obviously, the
relative intensity of the three components is not arbitrary. In Fig.
\ref{fig7}, the $\Pi$ PESs contribute to build up the maximum of the
line, whereas the the long blueshift tail arises from the $\Sigma$
PES.

No appreciable differences appear between $N=2000$ and 2500, the
largest drop we have calculated. This saturation has been also
observed in the experiment,\cite{Sti97,Sti99} although for larger
mean cluster sizes, about $N\sim 3000$. 
The peak energy is 79 cm$^{-1}$, a 10\% larger
than the experimental saturation value of 72 cm$^{-1}$,\cite{Sti97}
which indicates a fairly good agreement between theory and experiment.
The calculated width (FWHM) is $\sim 55$ cm$^{-1}$,
whereas the measured width for drops in the $N=1500-2000$ range is
$\sim 140-155$ cm$^{-1}$,\cite{Sti97} i.e., about three times
larger. We will show later on how thermal motion affects the
theoretical result.

Figure \ref{fig8} shows the total absorption spectrum of Ca attached
to $^4$He$_N$ droplets for several $N$ values. It is interesting to
notice the evolution of the absorption line as the number of atoms
increases. As a general rule, the smaller the drop, the smaller the
splitting of the $\Pi$ components. Indeed, they would be degenerate
if $N=1$, as Eq. (\ref{pairwise}) shows. This explains why the main
peak for $N=100$ is the narrower one, and is the reason why the main
peak is fairly apart from the $\Sigma$ shoulder. As $N$ increases,
so it does the splitting, while the three components of the peak
become broader. Eventually, if $N$ is large enough, it is not
possible to distinguish the components of the absorption line. It is
also obvious that line broadening due to effects not considered here
may wash out the blueshifted shoulder found for small $N$ values.

The inset  in Fig. \ref{fig8} shows the calculated shifts
relative to the gas-phase transition, compared with the experimental
values.\cite{Sti97} One can appreciate a small oscillation for the
largest drops; it is a genuine effect produced by the dimple structure.
Further insight can be gained from the study of the absorption
spectrum as a function of ${\cal Z}_0$.
To this end, we display in the top panel of Fig. \ref{fig9} 
the absorption spectra for the $N=500$ droplet corresponding to
${\cal Z}_0$ values from 16 to 19 \AA{} in 0.5 \AA{} steps, all of them
within the drop surface region 
(the equilibrium value is ${\cal Z}_0$= 17.45 \AA{}).
The inset in Fig. \ref{fig9} shows the
dependence of the relative shift on the location of the Ca atom.
One can see 
how a 3 \AA{} dispersion in the position of the impurity generates 
a $\sim 45$ cm$^{-1}$ change in the shift, showing in a quantitative way
the well known 
sensitivity of this quantity to the structure and depth of the dimple.
It is worth seeing how the shift decreases as the
distance between the centers of mass increases, and at the same time
the absorption peak becomes more asymmetric as the Ca environment
does (see also Subsection C). 

\subsection{Thermal broadening}

As we have indicated, for very attractive impurities, the foreign
atom is fully solvated, and its delocalization in the bulk of the
drop due to thermal motion hardly introduces a significant change in
the line shape. When the dopant is in a dimple state, it has to be
checked whether thermal motion  may have observable effects on the
absorption spectrum, as Fig. \ref{fig6} seems to indicate for
calcium.

To ascertain this effect, we have carried out a thermal average of
the spectrum using an approximate expression for the probability
density. Referring the energy $E_i$ of a given ${\cal Z}_{0_i}$
configuration to the equilibrium value, $\Delta E_i = E_i - E_{gs}$,
and neglecting the kinetic energy of the impurity and the displaced
fluid, we write the probability density for the position 
${\cal Z}_{0_i}$ of the Ca atom as
\begin{equation}
w_i=\frac{{\cal Z}^2_{0_i} e^{- \Delta E_i/k_B T} } {\Sigma_j \,
{\cal Z}^2_{0_j} e^{- \Delta E_j/k_B T} \Delta {\cal Z}_{0_j} } \; ,
\end{equation}
where $k_B$ is the Boltzmann constant, $T=0.4$ K, and the sum
-actually integral- runs on the selected ${\cal Z}_{0_i}$
configurations.\cite{Note}
The ${\cal Z}^2_{0_i}$ factor takes care of the
relative volume available to each configuration. With this
definition the probability of finding the Ca atom between ${\cal
Z}_{0_i}$ and ${\cal Z}_{0_i}+\Delta {\cal Z}_{0_i}$ is $w_i\Delta
{\cal Z}_{0_i}$. We show in the top panel of Fig. \ref{fig6} the
probability densities $w_i$ corresponding to the configurations
displayed in the bottom panel. We see that there is a non-negligible
probability of finding the Ca atom in a broad region of the
drop surface, and consequently we have addressed, as a case of study,
the statistical properties of the
Ca@$^4$He$_{500}$ system at this temperature. 

We have found that the mean
position, calculated as $\langle{\cal Z}_0\rangle = \Sigma_i
{\cal Z}_{0_i} w_i \Delta {\cal Z}_{0_i}$, and the standard
deviation, calculated as $\sigma({\cal Z}_0)=\sqrt{\langle{\cal
Z}_0^2\rangle-\langle{\cal Z}_0\rangle^2}$ are 17.38 \AA{} and 0.76
\AA, respectively.
This dispersion in the position generates a
dispersion in the value of the shift. To quantify this effect we
have evaluated the mean value of the shift, calculated as
$\langle\Delta\omega\rangle = \Sigma_i {\Delta\omega_i} w_i \Delta
{\cal Z}_{0_i}$, and its standard deviation, calculated as
$\sigma(\Delta\omega)=
\sqrt{\langle{\Delta\omega}^2\rangle-\langle{\Delta\omega}\rangle^2}$,
using the values shown in the inset of the top panel of Fig.
\ref{fig9}, which correspond to the seven lower energy
configurations in Fig. \ref{fig6}. We have obtained
$\langle\Delta\omega\rangle\pm\sigma(\Delta\omega)=$ 63.8 $\pm$ 11.5
cm$^{-1}$.

The thermally averaged Ca absorption spectrum is shown in the bottom
panel of Fig. \ref{fig9} (solid line), as well as that corresponding
to the equilibrium configuration (dashed line). To carry out the
average, we have used the $I_i(\omega)$ in the top
panel of Fig. \ref{fig9}, and averaged them as $I(\omega) =
\Sigma_i I_i(\omega) w_i \Delta {\cal Z}_{0_i}$. This procedure,
consistent with the Franck-Condon principle, assumes that absorption
proceeds instantaneously on any of the frozen drop-Ca configurations
characterized by a ${\cal Z}_{0_i}$ value.

We are led to conclude that the thermal motion effect is rather
small. It increases the FWHM by about 10\%, from $\sim 49.5$
cm$^{-1}$ to $\sim 55.0$ cm$^{-1}$, still a factor of three smaller
than the experimental value. From Fig. \ref{fig6}, we expect a
similar effect for the $N=1000$ drop, and likely for larger drops.

\subsection{Calcium atoms attached to vortex lines in $^4$He drops}

Since $^4$He is superfluid, it is quite natural to wonder
about the appearance and detection of quantized vortices in
droplets, see e.g. Refs. \onlinecite{Bar06,Leh03} and references
therein. Adapting an idea originally put forward by Close et
al.,\cite{Clo98} it has been proposed\cite{Anc03b} that Ca atoms should
be the dopant of choice to detect vortices by means of  microwave
spectroscopy experiments. The rationale of this proposal is that
Ca atoms are barely stable on the drop surface and become
solvated in its interior in the presence of a vortex line.\cite{Anc03b}
These conclusions were drawn from DFT calculations using Meyer's
Ca-He potential which, as shown in Fig. \ref{fig1}, is  slightly
more attractive than recent potentials. If this
scenario were plausible, one would not need the
microwave spectroscopy experiments suggested in Ref. \onlinecite{Anc03b}
to detect a vortex state in a Ca@$^4$He$_N$ drop: LIF spectroscopy could
do the job, given the
sizeable difference between the blueshifts of the absorption lines
when Ca has been drawn inside the drop by the vortex
(similar in value to the liquid helium blueshift),
and when it resides in a dimple state in vortex-free drops.\cite{Sti97}

This has prompted us to re-analyze the structure of a large
$N=1000$ drop hosting a calcium atom attached to
a vortex line along the symmetry axis. Within DFT,
a robust method to generate vortex configurations in liquid
helium is described in Ref. \onlinecite{Pi07}.
We adapt it here to the case of helium drops.
For a $n=1$ quantum circulation vortex line, we start the imaginary
time evolution to solve Eq. (\ref{eq2}) from the initial state

\begin{equation}
\Psi(\mathbf{r}) = \frac{\rho^{1/2}(\mathbf{r})}{\sqrt{x^2+ y^2}}
\, (x + \imath \,y)
\label{eq6}
\end{equation}
if $x$ and $y$ are nonvanishing, and zero otherwise, where
$\rho(\mathbf{r})$ is the vortex-free Ca@$^4$He$_{1000}$ helium density.
After the minimization procedure is converged, we have checked that
the obtained final configuration is indeed a $n=1$ vortex state.

Figure \ref{fig10} shows equidensity lines for the equilibrium
configuration of  Ca@$^4$He$_{1000}$ with and without
a vortex line along its symmetry axis.
It can be seen that the vortex line draws the impurity towards the bulk
of the droplet, but it still resides in a deeper surface dimple.

Figure \ref{fig11} shows the absorption spectrum
for the two configurations displayed in Fig. \ref{fig10}.
The effect of the presence of the vortex on the absorption spectrum
of calcium is twofold. On the one hand, the maximum of the
absorption peak is shifted towards the bulk value because of the deeper
dimple. On the other hand, the FWHM increases by about a factor
of two. The reason is the spreading of the Ca wave function within the
stretched $U_{Ca}(\mathbf{r})$ well, that allows the atom to ``probe''
a wider region in the excited PES, thus increasing the width.
Notice also the larger splitting of the peaks that form the line due
to the more anisotropic helium environment. This also contributes
to increasing the width of the absorption peak.
Unfortunately, the experimental absorption line is so
broad and asymmetric that the extra shift caused by the vortex is not
enough to displace the line to a region where it could be distinguishable
on top of the vortex-free absorption line.

\section{Summary}

Within density functional theory, we have carried out a detailed
study of the  absorption spectrum of calcium atoms attached to
$^4$He$_N$ drops in the vicinity of the $4s4p$ $^1$P$_1 \leftarrow
4s^2$ $^1$S$_0$ transition, finding a semi-quantitative agreement
with experiment. To this end, we have improved our previous
implementation of the DF method by incorporating the zero point
motion of the impurity, and have carried out {\it ab-initio}
calculations to obtain the excited $^1\Sigma$ and $^1\Pi$ Ca-He
potentials needed to obtain the potential energy surfaces.

We have studied the drop structure, finding that the
``interference'' between the density oscillations
of the helium moiety arising from its intrinsic structure  and
those arising from the presence of
the impurity plays a role in the determination of the position of
the impurity. This may be relevant for the solvation of alkaline 
earth atoms, especially for magnesium.\cite{Mel05}

In a case of study, we have systematically addressed the
dependence of the relative shift on the position of the impurity,
quantitatively assessing the relevance of a proper description of the
dimple to reproduce the experimental results. We have 
statistically taken into account the influence of the thermal motion of the
impurity on the absorption line, concluding that it only
increases the line width by a modest amount.

Finally, we have addressed the Ca absorption spectrum when the helium drop
hosts a vortex line, and conclude that absorption
spectroscopy experiments on these drops would be likely unable to
ascertain the presence of vortical states. In spite of this,
the study of vortex lines pinned by calcium atoms in
superfluid helium drops is interesting by itself, especially the
evaluation of the atomic shift caused by the presence
of a vortex line.

\section*{Acknowledgments}

We would like to thank Francesco Ancilotto and Kevin Lehmann
for useful comments and discussions.
This work has been performed under Grant No. FIS2005-01414 from DGI,
Spain (FEDER), and Grant 2005SGR00343 from Generalitat de Catalunya.
A.H. has been funded by the Project HPC-EUROPA
(RII3-CT-2003-506079), with the support of the European Community -
Research Infrastructure Action under the FP6 ``Structuring the
European Research Area'' Programme.

\pagebreak

\begin{figure}[f]
\centerline{\includegraphics[width=14cm,clip]{fig1.eps}} 
\caption{
(Color online) Some recent $X^1\Sigma$ Ca-He pair potentials: (1)
Ref. \onlinecite{Par01}; (2)  Ref. \onlinecite{Hin03}; (3)  Ref.
\onlinecite{Lov04}; (4)  Ref. \onlinecite{Meyer}. 
} 
\label{fig1}
\end{figure}

\pagebreak

\begin{figure}[f]
\centerline{\includegraphics[width=14cm,clip]{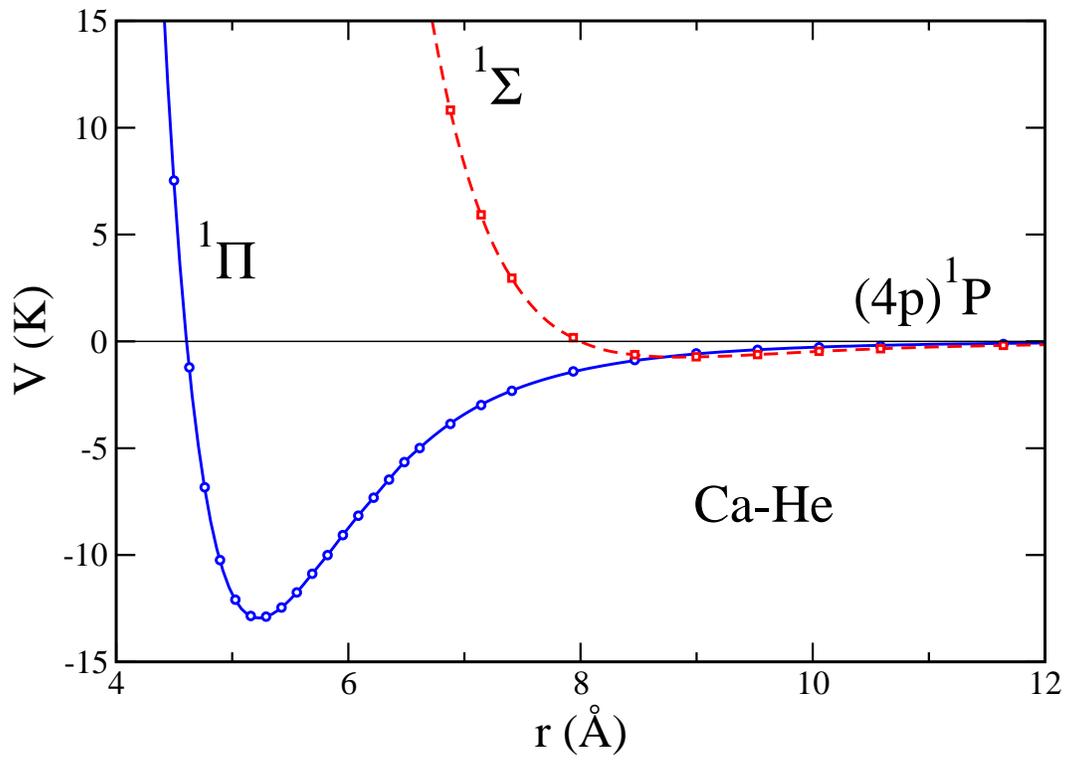}}
\caption{(Color online) 
Excited Ca-He pair potentials used in this work 
} 
\label{fig2}
\end{figure}

\pagebreak

\begin{figure}[f]
\centerline{\includegraphics[width=14cm,clip]{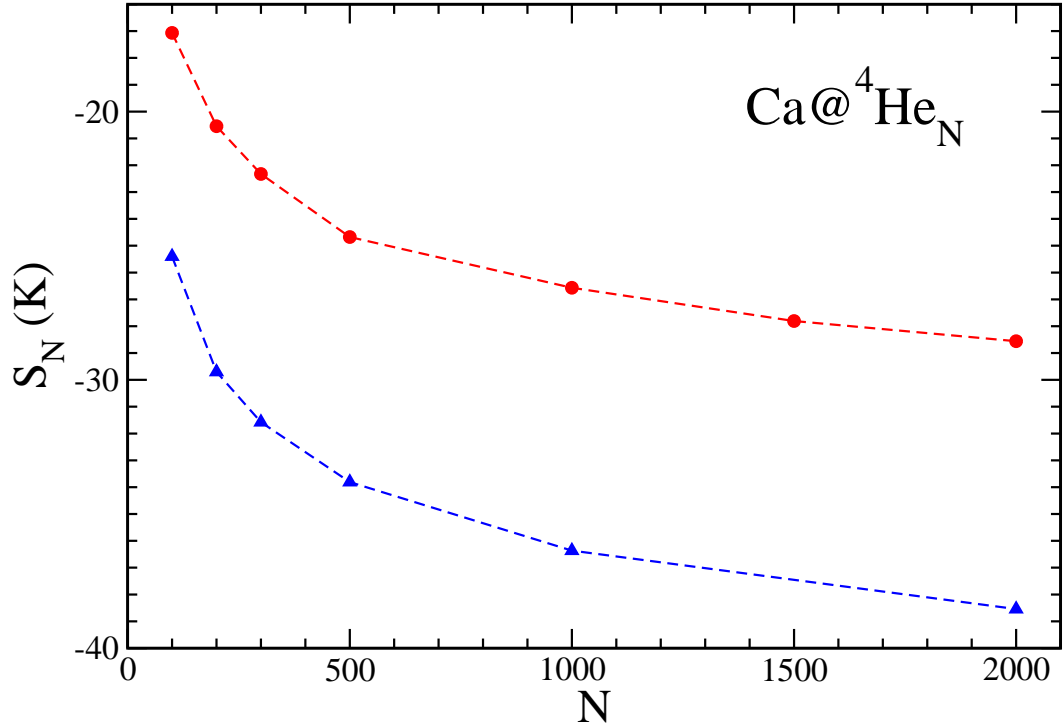}}
\caption{(Color online) 
Energy (K) of a calcium atom as a function
of the number of $^4$He atoms in the droplet (dots). Results
obtained treating calcium as an external field are also shown
(triangles).\cite{Her07} The lines have been drawn to guide the eye.
} 
\label{fig3}
\end{figure}

\pagebreak

\begin{figure}[f]
\centerline{\includegraphics[width=14cm,clip]{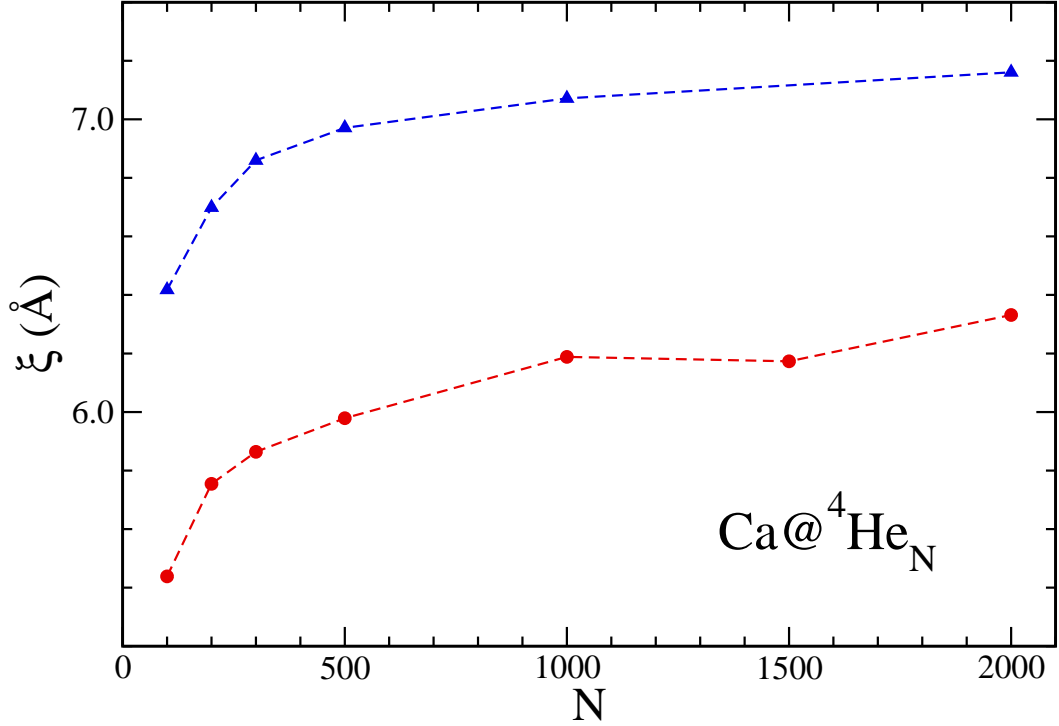}}
\caption{(Color online) 
Calcium dimples (\AA) as a function of the
number of $^4$He atoms in the droplet (dots). Results obtained
treating calcium as an external field\cite{Her07} are also shown
(triangles). The lines have been drawn to guide the eye. 
}
\label{fig4}
\end{figure}

\pagebreak

\begin{figure}[f]
\centerline{\includegraphics[width=14cm,clip]{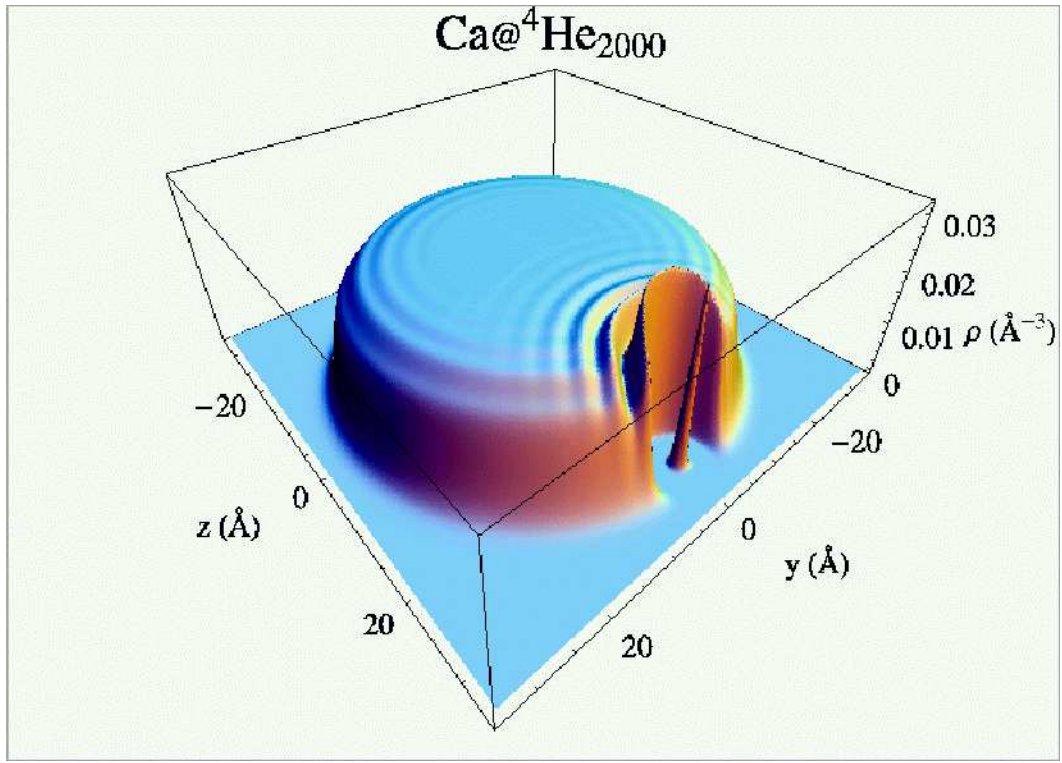}}
\caption{(Color online) Helium density of the Ca@$^4$He$_{2000}$
drop on the $x=0$ plane. The probability density of the calcium atom
is also shown, rescaled multiplying it by a $\rho_0$ factor for the
sake of clarity. 
} 
\label{fig5}
\end{figure}

\pagebreak

\begin{figure}[f]
\centerline{\includegraphics[width=12cm,clip]{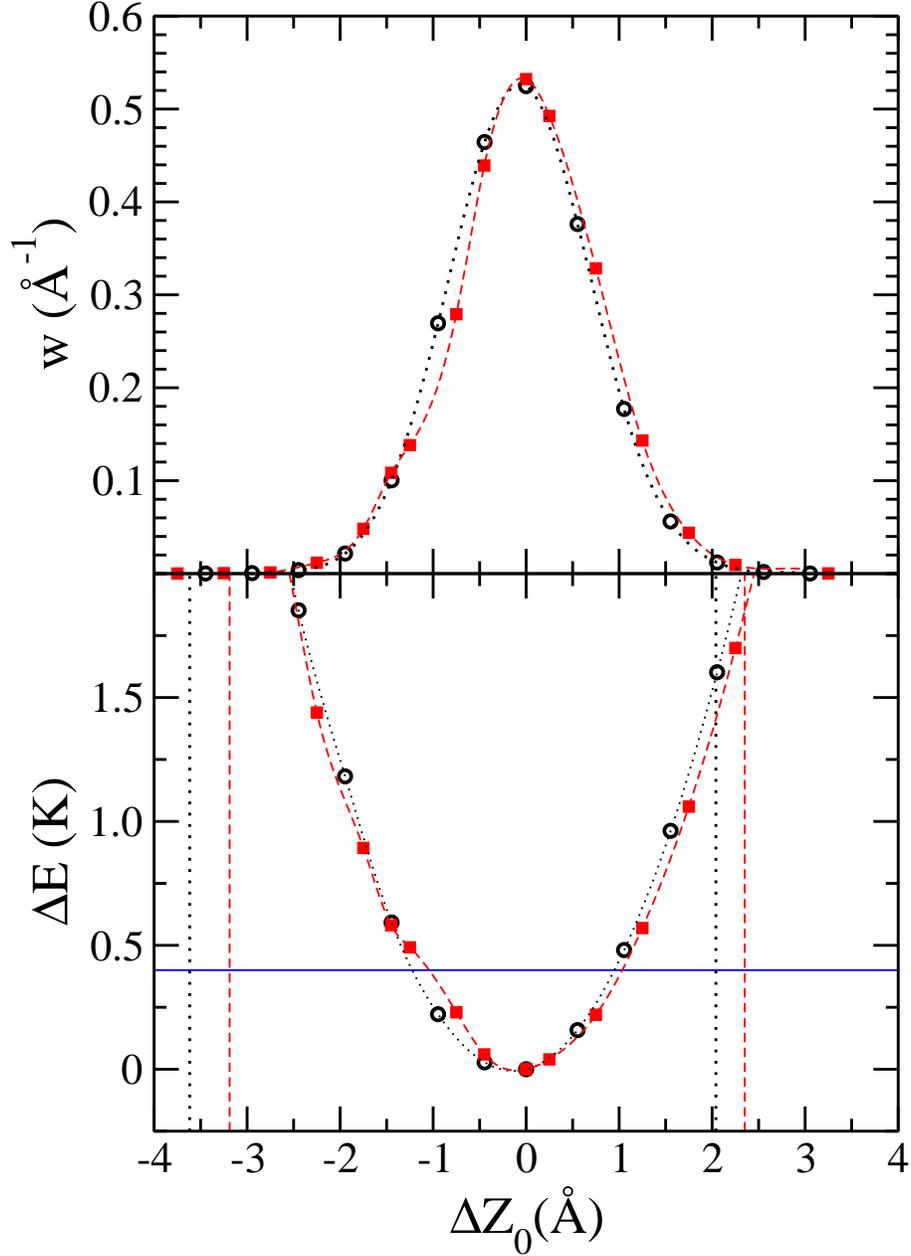}}
\caption{(Color online) 
Bottom panel: total energy (K) of
Ca@$^4$He$_{500}$ (circles) and Ca@$^4$He$_{1000}$ (squares) as a
function of ${\cal Z}_0$ (\AA), both referred to their equilibrium
values. The vertical lines delimit the drop surface
regions, and the horizontal line has been drawn 0.4 K above the
equilibrium energy. Top panel: probability densities for the
configurations displayed in the bottom panel. Circles correspond to
Ca@$^4$He$_{500}$, and squares to Ca@$^4$He$_{1000}$. In both
panels, dotted lines refer to the $N=500$ drop, and dashed lines to
the $N=1000$ drop; the lines (cubic splines) have been drawn to
guide the eye. 
} 
\label{fig6}
\end{figure}

\pagebreak

\begin{figure}[f]
\centerline{\includegraphics[width=14cm,clip]{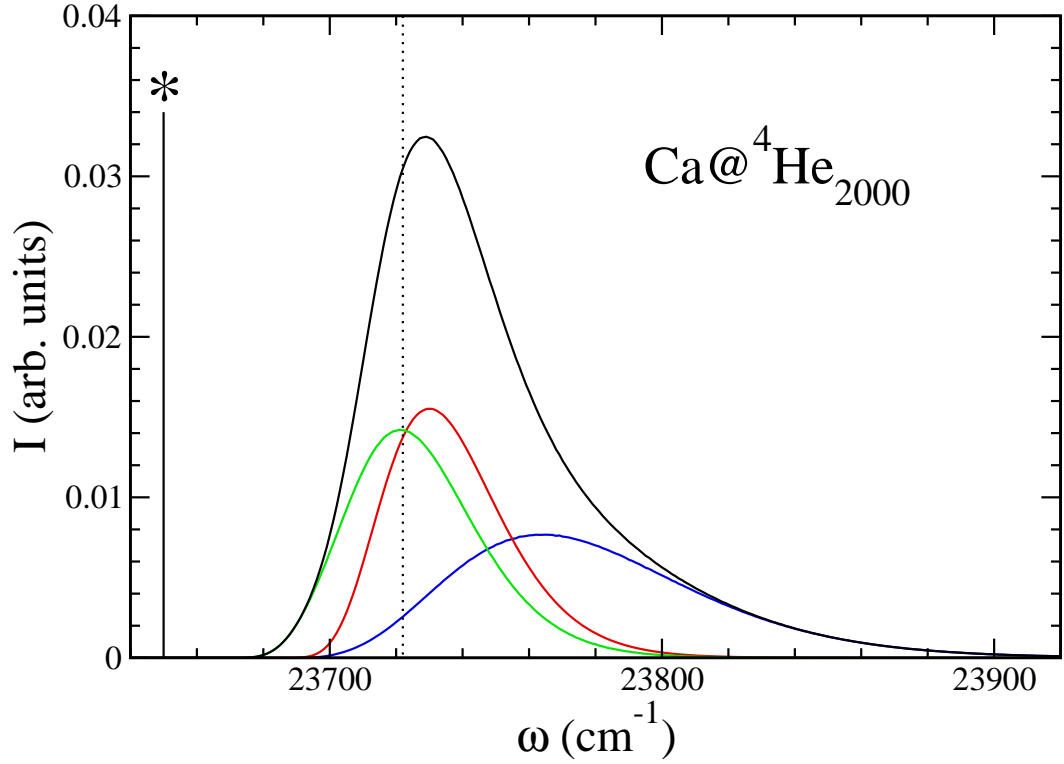}}
\caption{(Color online) 
Calcium absorption spectrum for $N=2000$ in
the vicinity of the $4s4p$ $^1$P$_1 \leftarrow 4s^2$ $^1$S$_0$
transition. The starred vertical line represents the gas-phase
transition, and the dotted vertical line  the experimental value
extracted from Fig. 3 of Ref. \onlinecite{Sti97}. The three
components of the absorption line, each arising from a different
excited PES, are also shown. 
} 
\label{fig7}
\end{figure}

\pagebreak

\begin{figure}[f]
\centerline{\includegraphics[width=14cm,clip]{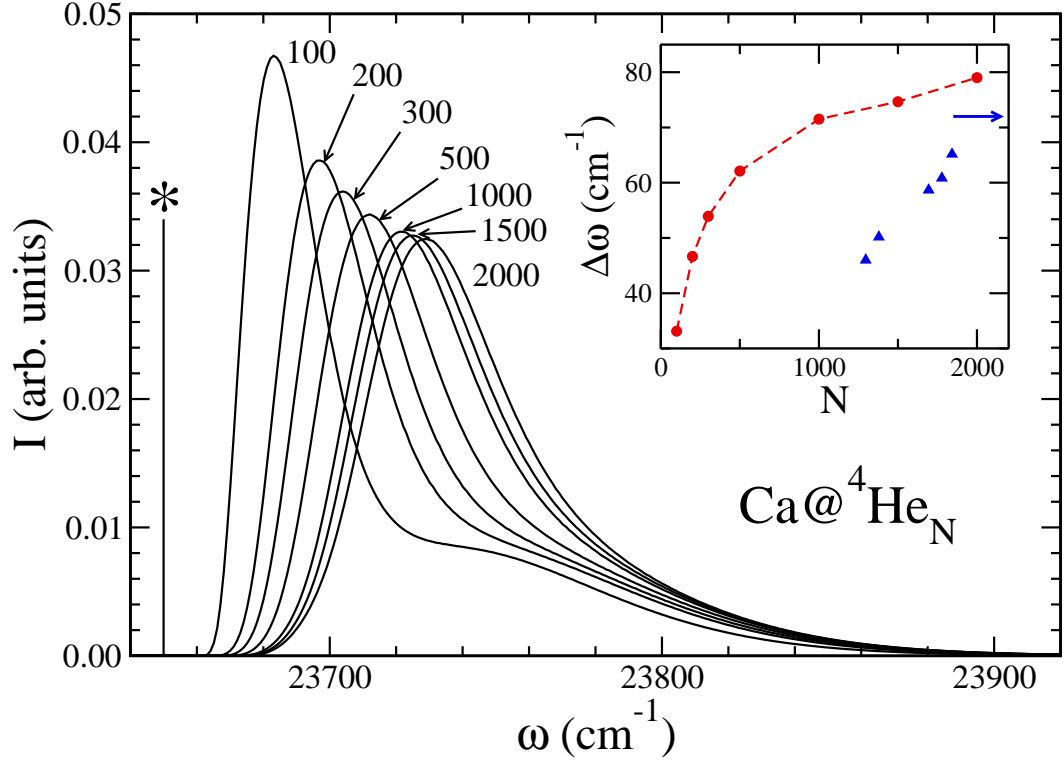}}
\caption{(Color online) 
Total absorption spectrum of Ca attached to
$^4$He$_N$ droplets in the vicinity of the $4s4p$ $^1$P$_1
\leftarrow 4s^2$ $^1$S$_0$ transition, for the indicated $N$ values.
The starred vertical line represents the gas-phase transition. The
inset shows the calculated shifts relative to the gas-phase
transition (dots), defined as  the energy of the maximum of the
absorption line minus the energy of the gas-phase transition. The
experimental values are also shown (triangles). The arrow indicates
the asymptotic experimental value.\cite{Sti97} 
} 
\label{fig8}
\end{figure}

\pagebreak

\begin{figure}[f]
\centerline{\includegraphics[width=14cm,clip]{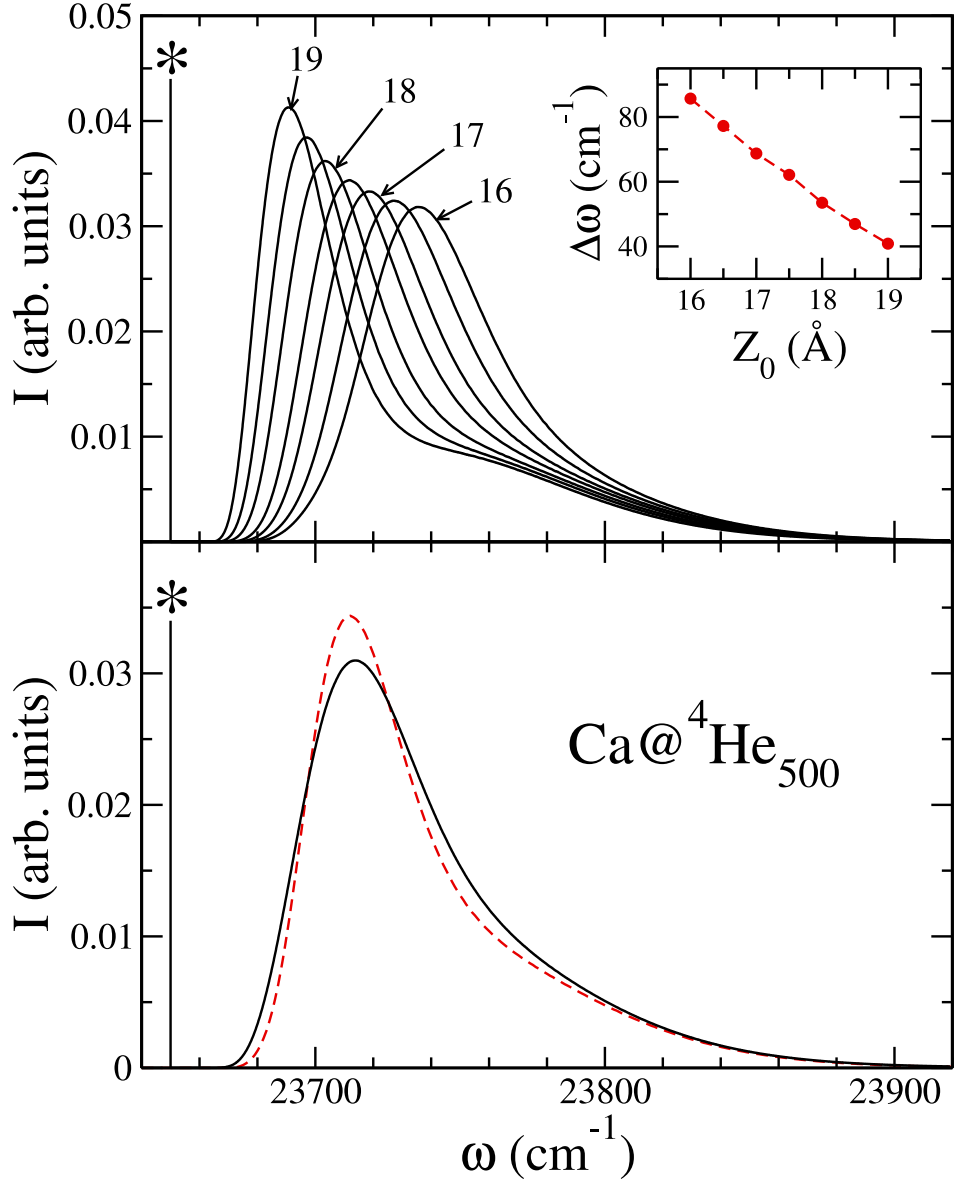}}
\caption{(Color online) 
Bottom panel: Thermally averaged absorption
spectrum of Ca in a $N=500$ drop (solid line); the dashed line
represents the absorption spectrum corresponding to the equilibrium
configuration. Top panel:  $I_i(\omega)$ spectra used to carry out
the average. The ${\cal Z}_{0_i}$ values go from 16 to 19 \AA {} in
0.5 \AA {} steps. The inset shows the associated atomic shifts. 
}
\label{fig9}
\end{figure}

\pagebreak

\begin{figure}[f]
\centerline{\includegraphics[width=9cm,clip]{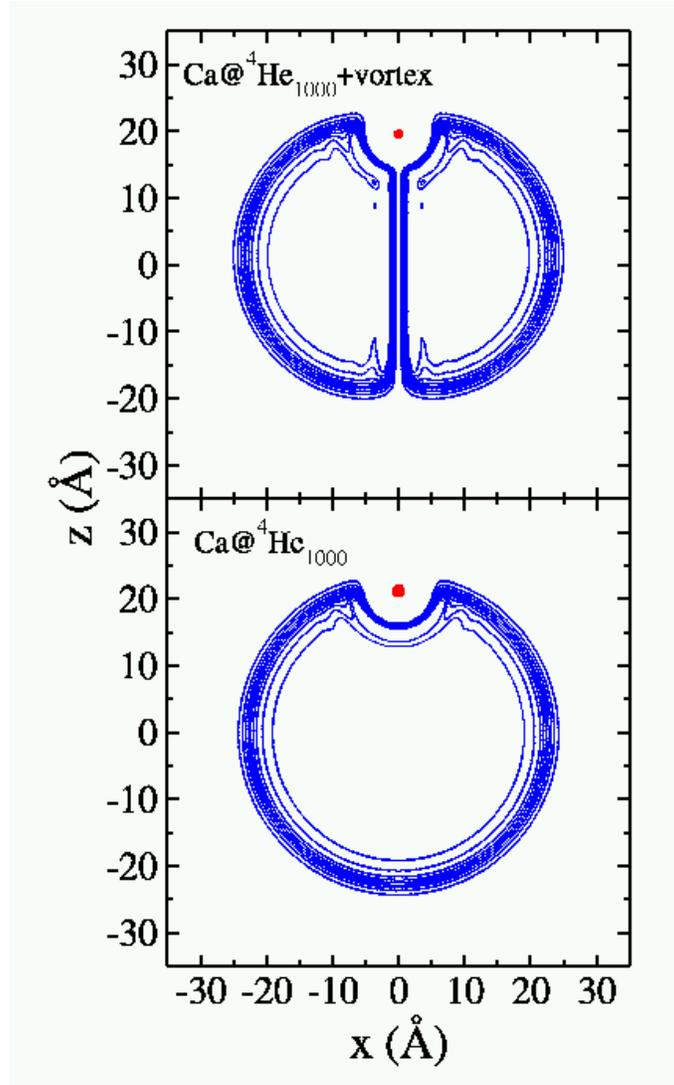}}
\caption{(Color online) 
Equidensity lines showing the equilibrium
configuration of a Ca atom on a $^4$He$_{1000}$ droplet with a
vortex line along its symmetry axis (top panel), and without it
(bottom panel).  The lines correspond to densities 0.9$\rho_0$ to
0.1$\rho_0$ in  0.1$\rho_0$ steps; these lines span the drop surface
region as well as the vortex core. The equidensity lines of the Ca
probability density are similarly plotted starting from its maximum
value. 
} 
\label{fig10}
\end{figure}

\pagebreak

\begin{figure}[f]
\centerline{\includegraphics[width=14cm,clip]{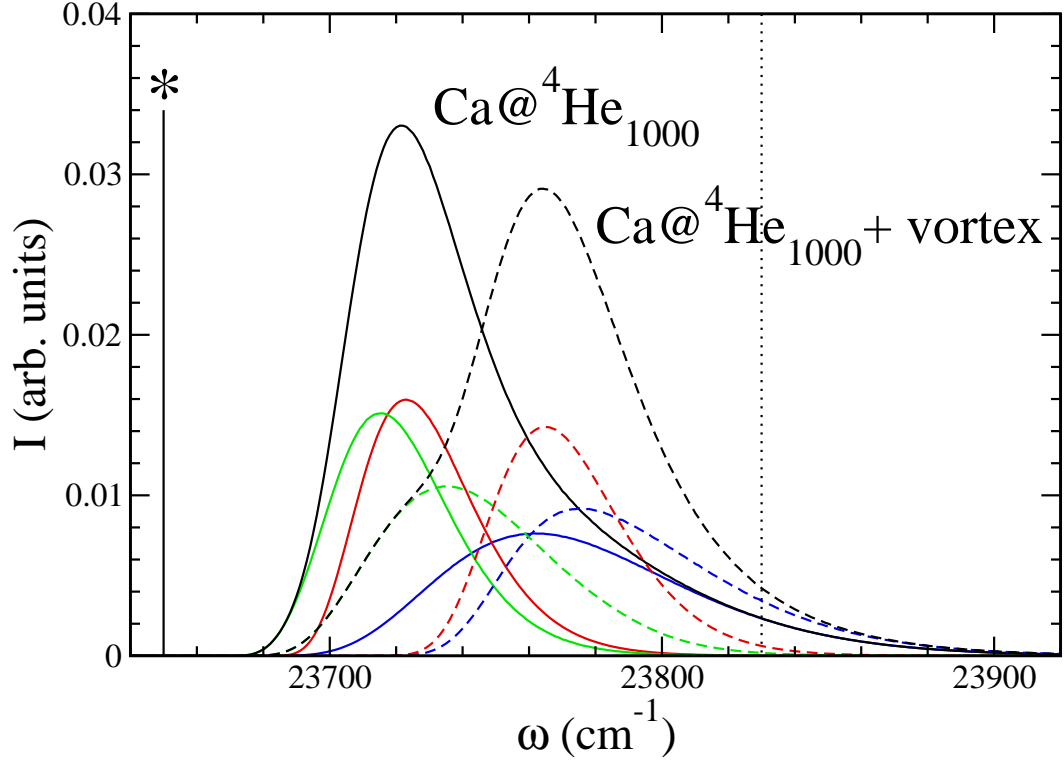}}
\caption{(Color online) 
Ca absorption spectrum for the $N=1000$
droplet of Fig. \ref{fig10} with (dashed lines) and without a vortex
line along its symmetry axis (solid lines). The absorption line has
been decomposed into its three components. The starred vertical line
represents the gas-phase transition, and the dotted vertical line
represents the experimental value for bulk liquid
$^4$He.\cite{Mor05} 
} 
\label{fig11}
\end{figure}


\begin{thebibliography}{99}
\bibitem{Note}
It is worth recalling that
the effective mass $m^*$ of the Ca atom will depend on $\mathcal{Z}_0$
since the volume of the displaced liquid depends on the size of the
dimple. We can obtain a lower bound for $m^*$ as the
mass of the free Ca atom, 40 au, and an upper bound as the
hydrodynamic mass of Ca in bulk helium,\cite{Leh02} 64 au.
In the harmonic approximation, we can estimate the quantum of
energy as
$\hbar\omega=\sqrt{\hbar^2E''({\cal Z}_0)|_{min}/m^*}$, finding for each 
bound a value around 0.3 K and 0.25 K, respectively.
In the same approximation, the population at $T=0.4$ K
of the excited states $p_{\text{ex}}$ relative to that
of the ground state $p_0$, calculated as
$p_{\text{ex}}/p_0=\sum_{n=1}^\infty\exp\left(-n\hbar\omega/k_BT\right)$,
is $\sim$ 0.3 and 0.4, respectively. These values are large enough
to justify the semiclassical approximation
for the average on the initial states, namely $\sum_n\exp\left(-\Delta
E_n/k_BT\right)\rightarrow 4 \pi \int d\mathbf{{\cal Z}_0}\,
{\cal Z}_0^2 \,\exp[-\Delta E({\cal Z}_0)/k_BT]$.
\end{thebibliography}
\end{document}